\documentstyle[12pt,fleqn]{article}
\topmargin - 0.8cm

\catcode`@=11
\@addtoreset{equation}{section}
\catcode`@=12
\mathchardef\SGamma="7100

\begin{document}
\title{\bf Effective equations of motion and initial conditions for
inflation in quantum cosmology}
\author{A.O. Barvinsky$^{1\dag}$ and  A.Yu. Kamenshchik$^{2\dagger}$}
\date{}
\maketitle
\hspace{-8mm}$^{1}${\em
Theory Department, Lebedev Physics Institute and Lebedev Research Center in
Physics, Leninsky Prospect 53,
Moscow 117924, Russia}\\
\hspace{-8mm}$^{2}${\em
L.D. Landau Institute for Theoretical Physics of
Russian Academy of \\ Sciences, Kosygina str. 2, Moscow
117334, Russia}
\begin{abstract}
We obtain effective equations of inflationary dynamics for the mean inflaton
and metric fields -- expectation values in the no-boundary
and tunneling quantum states of the Universe. The equations are derived in
the slow roll approximation taking the form of the local Schwiger-DeWitt
expansion. In this approximation effective equations follow from the
Euclidean effective action calculated on the DeSitter gravitational instanton
-- the basic element of the no-boundary and tunneling cosmological
wavefunctions. Effective equations are applied in the model of the inflaton
scalar field coupled to the GUT sector of matter fields and also
having a strong nonminimal coupling to the curvature. The inverse of its
big nonminimal coupling constant, $-\xi=|\xi|\gg 1$, serves as a
small parameter of the slow roll expansion and semiclassical expansion of
quantum gravitational effects. As a source of initial conditions for
effective equations we use a sharp probability peak recently obtained in the
one-loop approximation for the no-boundary and tunneling quantum states
and belonging (in virtue of big $|\xi|$) to the GUT energy scale much below
the Planck scale. Cosmological consequences of effective equations in the
tunneling quantum state predict a finite duration of inflationary stage
compatible with the observational status of inflation theory, whereas for
the no-boundary state they lead to the infinite inflationary epoch with
a constant inflaton field.
\end{abstract}
$^{\dag}$e-mail: barvin@td.lpi.ac.ru\\
$^{\dagger}$e-mail: kamen@landau.ac.ru\\

\section{Introduction}
\hspace{\parindent}
It has recently been shown that quantum cosmology with the no-boundary
\cite{HH,H,VilNB} and tunneling \cite{tun} quantum states of the Universe
can predict initial conditions for the inflationary scenario
\cite{qsi,qcr}. Until very recently this problem was regarded of marginal
significance, but now it becomes important in view of raising interest in
inflationary models with $\Omega\neq 1$ \cite{HawkTur,Lindop}. In particular,
the measure for the pre-inflationary initial conditions (prior probability)
is essential for finding the posterior probability of the present value of
$\Omega$.

Such an approach suffers from the known problems inherent in the
tree-level approximation of quantum cosmology -- the lack of normalizability
of the cosmological wavefunction and the absence of necessary probability
maxima \cite{Page,Vilenkin}. These problems can be resolved by including
the loop effects \cite{norm,BarvU,tunnel,tvsnb}.
They modify the distribution function of the quantum ensemble of inflationary
models with different initial values of the inflaton $\varphi$ -- a scalar
field driving the chaotic inflation with the Hubble constant $H=H(\varphi)$
\cite{inflation}. In the one-loop approximation the distribution function
of this field at the beginning of the Lorentzian quasi-DeSitter evolution
has the form \cite{norm,BarvU,tunnel,tvsnb}
	\begin{equation}
        \rho_{\rm N\!B,T}(\varphi)= {\rm const}\,
         e^{{}^{{\textstyle
	\mp{\mbox{\boldmath $I$}}(\varphi)-
        \mbox{\boldmath $\Gamma$}^{\rm 1-loop}(\varphi)}}}.   \label{1.1}
	\end{equation}
Here ${\mbox{\boldmath $I$}}(\varphi)$ is the classical Euclidean action of
the model at the gravitational instanton -- 4-dimensional (quasi)sphere
of the radius $1/H(\varphi)$ and
$\mbox{\boldmath $\Gamma$}^{\rm 1-loop}(\varphi)$ is the Euclidean
one-loop effective action of {\it all} quantum fields of the model calculated
at this instanton. Important peculiarity of this algorithm is that in
contrast with opposite signs of the tree-level part (minus and plus
correspond respectively to the no-boundary and tunneling quantum states)
the one-loop corrections are the same for both cosmological wavefunctions
\cite{tvsnb}.

Depending on the anomalous scaling behaviour of the particle physics model,
the one-loop term of (\ref{1.1}) can suppress big values of $\varphi$
making $\rho_{\rm NB,T}(\varphi)$ normalizable in the high energy limit
\cite{norm}. Moreover, in the model with large nonminimal coupling of
the inflaton field to curvature and typical couplings to Higgs, vector
gauge and spinor matter fields the distribution function (\ref{1.1}) has
a sharp peak at the grand unification energy scale. For the tunneling
quantum state this peak generates the inflationary scenario compatible
with the observational status of the inflation theory \cite{qsi,qcr}. In
particular, at least in context of closed cosmology considered in
\cite{qsi,qcr}, it is capable of producing sufficient but finite e-folding
number and thus gives rise to intermediate values of $\Omega\neq 1$ without
invoking the anthropic considerations of \cite{HawkTur} or using exotic
supergravity induced inflaton potentials of \cite{Lindop}.

These conclusions have been drawn
\cite{norm,BarvU,tunnel,qsi,qcr} in the assumption that the
solutions of {\it classical} equations of inflationary dynamics are
weighted by the quantum distribution function (\ref{1.1}). However,
the quantum effects qualitatively change the behaviour of the tree-level
distribution and, therefore, they are not small. This means that
within the same accuracy classical equations of motion should be replaced
by the {\it effective} equations for the {\it mean} inflaton field. Thus,
the main goal of this paper will be to derive such equations and infer
their consequences at the initial stage of inflation. In particular,
we will clarify the most important qualitative aspect of effective
dynamics -- the direction of evolution from the point of the probability
maximum in both quantum states of the Universe.

We consider the cosmological model with the total Lagrangian
	\begin{eqnarray}
	&&{\mbox{\boldmath $L$}}(g_{\mu\nu},\varphi,\chi,A_\mu,\psi)=
	{\mbox{\boldmath $L$}}(g_{\mu\nu},\varphi)
	+g^{1/2}\left(-\frac 12 \sum_\chi (\nabla\chi)^2
	-\frac 14 \sum_A F_{\mu\nu}^2(A)
	-\sum_\psi \bar{\psi}\hat{\nabla}\psi\right)\nonumber\\
	&&\qquad\qquad\qquad\qquad\qquad\qquad
	+{\mbox{\boldmath $L$}}_{\rm int}
	(\varphi,\chi,A_\mu,\psi)                       \label{1.2}
	\end{eqnarray}
containing the graviton-inflaton sector
	\begin{equation}
	{\mbox{\boldmath $L$}}(g_{\mu\nu},\varphi)
	=g^{1/2}\left\{\frac{m_{P}^{2}}{16\pi} R(g_{\mu\nu})
	-\frac{1}{2}\xi\varphi^{2}R(g_{\mu\nu})
	-\frac{1}{2}(\nabla\varphi)^{2}
	-\frac{1}{2}m^{2}\varphi^{2}
	-\frac{\lambda}{4}\varphi^{4}\right\},      \label{1.3}
	\end{equation}
with a big negative nonminimal coupling constant $-\xi=|\xi|\gg 1$, and
generic GUT sector of Higgs $\chi$, vector gauge $A_\mu$ and spinor
fields $\psi$ coupled to the inflaton via the interaction term
	\begin{eqnarray}
	{\mbox{\boldmath $L$}}_{\rm int}
	=\sum_{\chi}\frac{\lambda_{\chi}}4
	\chi^2\varphi^2
	+\sum_{A}\frac12 g_{A}^2A_{\mu}^2\varphi^2+
        \sum_{\psi}f_{\psi}\varphi\bar\psi\psi
	+{\rm derivative\,\,coupling}.             \label{1.4}
	\end{eqnarray}

This model is of a particular interest for a number of reasons. Firstly,
from the phenomenological viewpoint a strong nonminimal coupling allows
one to solve the problem of exceedingly small $\lambda$ (because the
observable magnitude of anisotropy $\Delta T/T\sim 10^{-5}$ is proportional
in this model to the ratio $\sqrt{\lambda}/|\xi|$ \cite{nonmin}), and for
positive $\xi$ \cite{GBLinde} this model is useful for generating inflation
with $\Omega\neq 1$ \cite{LindeM}. Secondly, this coupling is inevitable
from the viewpoint of renormalization theory. Finally, for a wide class of
GUT-type particle physics models (\ref{1.2}) due to a big value $|\xi|$
there exists a sharp probability peak in $\rho_{\rm NB,T}(\varphi)$
\cite{qsi,qcr}. This peak belongs to GUT energy scale -- a characteristic
value of the effective Hubble constant driving inflation, which is
proportional to $m_P\sqrt{\lambda}/|\xi|\sim 10^{-5}m_P$. This, in its
turn, justifies the use of GUT for matter field sector of the model, because
this energy scale is much below the supersymmety and string theory scales.

In the Lagrangian (\ref{1.2}) the inflaton field can be regarded as a
component of one of the Higgs multiplets $\chi$, which has a nonvanishing
expectation value in the cosmological quantum state. The inflaton has a
quartic selfinteraction and mass term $m^2\varphi^2/2$ which for generality
can be negative $(m^2<0)$, thus, including the case of symmetry breaking.
The choice of the interaction Lagrangian (\ref{1.4}) is dictated by the
renormalizability of the matter field
sector of the theory (\ref{1.2}) and by the requirement of local
gauge invariance with respect to arbitrary Yang-Mills group of vector
fields $A_\mu$. The terms of derivative coupling in (\ref{1.4}) should be
chosen to guarantee the latter property, but their form is not important.
On the contrary, as shown in \cite{qsi,qcr}, the quantum gravitational
effects generating the probability peak of the above type crucially depend
on the nonderivative part of the interaction Lagrangian.

The organization of the paper is as follows. In Sect.2 we consider
the classical inflation dynamics of the graviton-inflaton model (\ref{1.3})
depending on initial conditions generated by the probability peak of
$\rho_{\rm NB,T}(\varphi)$. In Sect.3 the derivation of effective
equations for mean fields is outlined on the basis of the Euclidean effective
action of the theory. Sects.4 and 5 are devoted to the effective action
calculations in the slow-roll approximation equivalent in this context to
the local Schwinger-DeWitt expansion. In Sect.6 we compare the cosmological
consequences of the obtained effective equations for the tunneling and
no-boundary quantum states and conclude that phenomenologically the
tunneling wavefunction is a more preferable candidate for the initial state
of the early Universe. In concluding section we briefly comment on the
possibility of extending our results to the case of the open inflation
originating from the Hawking-Turok instanton via the no-boundary proposal
\cite{HawkTur} and the tunnelling proposal of Linde \cite{Lindop}. We also
discuss the limitations of the obtained results and their nonperturbative
extension and conjecture on the contribution of the quantum mechanical sector
of the model (quantum homogeneous mode of the inflaton field) which goes
beyond the scope of this paper and will be considered in future publications.

\section{Classical equations of motion for chaotic inflation}
\hspace{\parindent}
Here we consider classical equations of inflationary dynamics in the model
with the Lagrangian (\ref{1.3}). For our future purposes we generalize
this model to have generic coefficients -- functions of the inflaton field --
for the effective cosmological term $V(\varphi)$ and effective gravitational
constant $U(\varphi)$ \cite{renorm,khalat}:
	\begin{equation}
	S[g_{\mu\nu},\varphi]=\int d^{4}x\, g^{1/2}
	\left(U(\varphi)\,R(g_{\mu\nu})-
	\frac12(\nabla\varphi)^2
	-V(\varphi)\right).         \label{action1}
	\end{equation}
With the minisuperspace Robertson-Walker ansatz for the spacetime metric
of spatially closed cosmological model (in the cosmic time gauge, $g_{00}=-1$,
and with the scale factor $a=a(t)$) with the spatially homogeneous
inflaton field $\varphi=\varphi(t)$, the equations of motion take the form:
	\begin{eqnarray}
	&&12 a U\ddot{a}+6U\dot{a}^{2}+12 a
	U'\dot{a}\dot{\varphi}
	+6 a^{2}U''\dot{\varphi}^{2}     \nonumber \\
	&&\qquad\qquad\qquad\qquad\qquad
	+6a^{2} U'\ddot{\varphi}
	+6U+\frac{3}{2}a^{2}\dot{\varphi}^{2}
	-3a^{2}V=0,                                \label{equation1}\\
	&&a^{3}\ddot{\varphi}+3a^2\dot{a}\dot{\varphi}
	-6aU'\dot{a}^{2}
	-6a^{2}U'\ddot{a}-6aU'+a^{3}V'=0,          \label{equation2}\\
	&&a^{3}\left(V+\frac{\dot{\varphi}^{2}}{2}\right)-
	6a^3 U\left(\frac{\dot{a}^2}{a^2}+\frac 1{a^2}\right)
	-6a^{2}U'\dot{a}\dot{\varphi}
	=0.                                        \label{constraint}
	\end{eqnarray}
where the dots denote time derivatives and the prime denotes the derivative
of the coefficient functions of the Lagrangian with respect to $\varphi$.
The dynamical equations (\ref{equation1}) - (\ref{equation2}) can be solved
with respect to second order time derivatives of $a$ and $\varphi$ with the
result for $\ddot{\varphi}$:
	\begin{eqnarray}
	&&\ddot{\varphi} = \frac{1}{U+3U'^{2}}\,\left(
	-\frac{3U\dot{a}\dot{\varphi}}{a}
	-\frac{6U'^{2}\dot{a}\dot{\varphi}}{a}
	+\frac{3UU'\dot{a}^{2}}{a^{2}}\right.\nonumber \\
	&&\qquad\qquad\qquad\left.-3U'U''\dot{\varphi}^{2}
	-\frac{3U'\dot{\varphi}^{2}}{4}+\frac{3UU'}{a^{2}}
	+\frac{3VU'}{2}-UV'\right).                  \label{equation3}
	\end{eqnarray}
The constraint equation (\ref{constraint}) in its turn can be solved with
respect to the Hubble ``constant'' $\dot{a}/a$ which, when substituted to
(\ref{equation3}), gives
	\begin{equation}
	\ddot{\varphi}+3\,\frac{\dot{a}}{a}\,\dot{\varphi}
	-\frac{1}{U+3U'^{2}}
	\left(2VU'-UV'-\frac12\,U'\dot{\varphi}^{2}
	-3U'U''\dot{\varphi}^{2}\right)=0.    \label{equation4}
	\end{equation}

In the slow-roll regime one can neglect the terms quadratic (and of higher
powers) in time derivatives of $\varphi$, so that the system of
equations reduces to the expression for the effective Hubble constant
	\begin{eqnarray}
	\left(\frac{\dot a}a\right)^2=H^2(\varphi)\simeq
	\frac{V(\varphi)}{6U(\varphi)}               \label{equation5a}
	\end{eqnarray}
and the equation of motion for the inflaton field
	\begin{eqnarray}
	&&\ddot{\varphi}+3H\dot{\varphi}-F(\varphi)=0,  \label{equation5b}\\
	&&F(\varphi)\simeq\frac{2VU'-UV'}{U+3U'^{2}}=
	-\frac{U^3}{U+3U'^2}\,
	\frac d{d\varphi}\left(\frac V{U^2}\right),    \label{equation5}
	\end{eqnarray}
evolving under the action of the {\it rolling} force $F(\varphi)$ and the
``friction'' force $-3H\dot{\varphi}$. For constant $U\equiv m_P^2/16\pi$
the rolling force reduces to the usual gradient of the scalar field potential,
while for a nonminimal inflaton it is proportional to the gradient of the
modified potential $V(\varphi)/U^2(\varphi)$ renormalized by the nonminimal
coupling\footnote
{
The combination $V(\varphi)/U^2(\varphi)$ coincides with the inflaton
potential in the Einstein frame of the action (\ref{action1}) that can be
obtained by the conformal transformation of the metric and special
reparametrization of the inflaton field \cite{renorm}.
}.

Quantum initial conditions for inflation stage crucially depend
on the value of the parameter
	\begin{eqnarray}
	\delta=-\frac{8\pi\,|\xi|\,m^2}{\lambda\,m_P^2}    \label{delta}
	\end{eqnarray}
characterizing the model (\ref{1.3}). As shown in \cite{qsi,qcr} the
probability peak in the distribution functions of $\varphi$ for the
no-boundary and tunneling quantum states exists in complimentary domains
of $\delta$: for $\delta<-1$ in the no-boundary case and for $\delta>-1$
for the tunneling one. The parameters of this peak -- mean value and
relative width -- for a large value of the nonminimal coupling $|\xi|\gg 1$
are given in both cases by the same expressions
	\begin{eqnarray}
	&&\varphi_I= m_{P}\sqrt{\frac{8\pi|1+\delta|}{|\xi|
	{\mbox{\boldmath$A$}}}},\,\,\,\,\,
	H(\varphi_I)=
        m_{P}\frac{\sqrt{\lambda}}{|\xi|}
	\sqrt{\frac{2\pi|1+\delta|}
	{3{\mbox{\boldmath $A$}}^2}},             \label{2.1} \\
	&&\frac{\Delta\varphi}{\varphi_I}\sim
	\frac{\Delta H}{H}\sim
	\frac 1{\sqrt{12{\mbox{\boldmath $A$}}}}
        \frac{\sqrt{\lambda}}{|\xi|},              \label{peak}
	\end{eqnarray}
where $\mbox{\boldmath$A$}$ is the following combination of Higgs, vector
gauge boson and Yukawa coupling constants of the GUT-inflaton interaction
Lagrangian (\ref{1.4})
	\begin{eqnarray}
	{\mbox{\boldmath $A$}} = \frac{1}{2\lambda}
	\left(\sum_{\chi} \lambda_{\chi}^{2}
	+ 16 \sum_{A} g_{A}^{4} - 16
	\sum_{\psi} f_{\psi}^{4}\right),   \label{A}
	\end{eqnarray}
but correspond to opposite signs of the rolling force generating different
inflationary scenarios
	\begin{eqnarray}
	F(\varphi)=-\frac{\lambda m_P^2
	(1+\delta)}{48\pi\xi^2}\,\varphi+O(1/|\xi|^3).    \label{2.2}
	\end{eqnarray}
For no-boundary state the maximum of the
distribution function $\varphi_I$ belongs to the negative slope of the
potential $V(\varphi)/U^2(\varphi)$ and $F(\varphi_I)>0$, $1+\delta<0$.
This results in the slow-roll regime in the infinitely long inflationary
stage with ever growing inflaton field
	\begin{eqnarray}
	\dot{\varphi}\simeq\frac 1{3H(\varphi)}\,F(\varphi).  \label{2.10}
	\end{eqnarray}
For the tunneling state $\varphi_I$ lies on the positive slope of
$V(\varphi)/U^2(\varphi)$ with $F(\varphi_I)<0,\;1+\delta>0,$ and the
inflationary stage has a slowly decreasing scalar field and finite
duration with an approximate e-folding number \cite{qsi,qcr}\footnote
{
Eq. (\ref{2.11}) is valid up to numerical factor
$1+O(\varepsilon\ln\varepsilon),\,\varepsilon=\mbox{\boldmath$A$}/32\pi^2$,
slightly different from unity. These corrections correspond to late stages
of inflation when lower order terms of $U(\varphi)$ and $V(\varphi)$ come
into game. Unfortunately, the references \cite{qsi,qcr,tvsnb} contain a typo
in the leading term of the e-folding number (\ref{2.11}): $8\pi^2$ instead
of the correct value $48\pi^2$. This correction raises the upper bound
on the universal combination of coupling constants (following from the
lower bound on $N$, $N\geq 60$), $\mbox{\boldmath$A$}\leq 7.9$, but does
not change qualitatively the conclusions of \cite{qsi,qcr,tvsnb} leaving us
with a small parameter $\mbox{\boldmath$A$}/32\pi^2\ll 1$.
}
	\begin{eqnarray}
	N\simeq -\int\limits_0^{\varphi_I}
	d\varphi\,\frac{H(\varphi)}{\dot\varphi}
	\simeq \frac{48\pi^2}{\mbox{\boldmath $A$}}.   \label{2.11}
	\end{eqnarray}

These conclusions are, however, based on classical equations of motion in
contrast with the quantum nature of initial conditions originating from
the maximum of the quantum distribution function -- the quantity drastically
different from its tree-level counterpart. The purpose of this
paper is to cure this mismatch by replacing the classical equations with
the effective equations for expectation values.

\section{Effective equations for expectation values}
\hspace{\parindent}
Effective equations of motion for expectation values of operators of
the total system of fields
	\begin{eqnarray}
	&&\phi(x)=\big<{\mbox{\boldmath $\Psi$}}|\hat{\phi}(x)
	|{\mbox{\boldmath $\Psi$}}\big>,      \label{3.1}\\
	&&\hat{\phi}(x)=\hat{\varphi}(x),\hat{\chi}(x),
	\hat{\psi}(x),\hat{A}_\mu(x),\hat{g}_{\mu\nu}(x),... \label{3.2}
	\end{eqnarray}
with respect to any quantum state including the no-boundary and tunneling
ones,
	$|{\mbox{\boldmath $\Psi$}}\big>=
	|{\mbox{\boldmath $\Psi$}}\big>_{\rm NB},\,
	|{\mbox{\boldmath $\Psi$}}\big>_{\rm T}$,
have a generic form
	\begin{eqnarray}
	\frac{\delta S[\,\phi\,]}{\delta\phi(x)}
	+J^{\rm rad}(x)=0.                              \label{3.3}
	\end{eqnarray}
Here the radiation current $J^{\rm rad}(x)$ accumulates all quantum
corrections which begin with the one-loop contribution
	\begin{eqnarray}
	J^{\rm rad}(x)=\frac 1{2i}\int dy\,dz\,
	\frac{\delta^3 S[\,\phi\,]}
	{\delta\phi(x)\,\delta\phi(y)\,\delta\phi(z)}
	\,G(z,y)+...                                     \label{3.4}
	\end{eqnarray}
containing the Wightman function of quantum disturbances in a given
quantum state
	\begin{eqnarray}
	G(z,y)=\big<{\mbox{\boldmath $\Psi$}}|\,\Delta\hat{\phi}(z)
	\,\Delta\hat{\phi}(y)\,
	|{\mbox{\boldmath $\Psi$}}\big>,\,\,\,
	\Delta\hat{\phi}(y)\equiv\hat{\phi}(y)-\phi(y).  \label{3.5}
	\end{eqnarray}
In the sector of spacetime metric $\phi(x)=g_{\mu\nu}(x)$ this radiation
current coincides, in particular, with the expectation value of the
quantum matter stress tensor $J^{\mu\nu\, \rm rad}(x)=
\big<{\mbox{\boldmath $\Psi$}}|\hat{T}^{\mu\nu}(x)
|{\mbox{\boldmath $\Psi$}}\big>$.

The calculation of $J^{\rm rad}(x)$ even in the one-loop approximation
generally presents a hard problem, because the Wightman Green's function
in the external mean field of arbitrary configuration comprises a very
complicated nonlocal object that cannot be obtained exactly. Fortunately,
the model in question has a number of peculiarities that essentially
simplify calculations and look as follows.

To begin with, note that in our model with a large negative constant
$-\xi=|\xi|\gg 1$ and slowly varying inflaton field the nonminimal
coupling efficiently implies a replacement of the Planckian mass
parameter $m_P^2$ by the effective mass of a much bigger magnitude
	\begin{eqnarray}
	m_P^2\rightarrow m_{\rm eff}^2=
	m_P^2+8\pi|\xi|\varphi^2\gg m_P^2.             \label{3.6}
	\end{eqnarray}
This essentially improves the semiclassical expansion of quantum
gravitational effects, because this expansion goes in inverse powers of
$m_{\rm eff}^2$ rather than of $m_P^2$.

Big value of $|\xi|$ has also another important effect caused by the Higgs
mechanism for all matter fields interacting with inflaton. As discussed in
\cite{qsi,qcr}, due to this interaction the corresponding matter particles
acquire masses proportional to the background value of the inflaton field,
$m_{\rm part}^2\sim\varphi^2$, but in view of eq.(\ref{equation5a}) (with
$V$ and $U$ read off the classical Lagrangian (\ref{1.3})) the spacetime
curvature has an order of magnitude
$R\sim H^2\sim \lambda\varphi^2/|\xi|\ll\varphi^2$.
Therefore, quantum contribution of matter fields can be expanded in local
Schwinger-DeWitt series \cite{DW,BarvV} in powers of the curvature to mass
squared ratio
	\begin{eqnarray}
	\frac{R}{m_{\rm part}^2}
	\sim\frac\lambda{|\xi|}\ll 1,               \label{3.7}
	\end{eqnarray}
the first few terms giving a dominant contribution polynomial in
$|\xi|\gg 1$. In the limit of big $|\xi|$ these terms dominate over
contribution of all other fields uncoupled to inflaton and, in particular,
over the contribution of the graviton-inflaton sector. Below we show this
property by direct calculations in the one-loop approximation. The
mechanism of this result is based on the improvement of semiclassical
expansion due to the replacement (\ref{3.6}) and, apparently, holds in
multi-loop orders as well.

Finally, in our setting of the problem only two fields have nonvanishing
expectation values -- spacetime metric and inflaton scalar field. We assume
that the slow roll approximation remains applicable also at the quantum level,
which means that in the leading order of this approximation the mean
spacetime metric corresponds to DeSitter geometry and the mean inflaton field
is a spacetime constant scalar. It is also well known that for all
{\it massive and/or spatially inhomogeneous} modes of fields the no-boundary
and tunneling cosmological states turn out to be the Euclidean DeSitter
invariant vacuum \cite{Laf,VilVach}. Therefore, in the one-loop approximation the
Wightman Green's function (\ref{3.5}) of such modes $\Delta\phi(y)$ --
solutions of linerized Heisenberg operators -- is uniquely fixed by the
choice of this vacuum \cite{Allen}. The exception from this simple rule are
massless scalar fields for which the Euclidean DeSitter vacuum does not exist
\cite{Allen} and {\it effectively} massless inflaton mode of the
graviton-inflaton sector of the model\footnote
{
The linearized equation of the inflaton mode has a very small mass parameter
suppressed by the factor $O(\lambda/\xi^2)$, and its smallness guarantees
the validity of the slow roll approximation -- the corner stone of inflation
theory.
}.
The quantum state of this mode is not the DeSitter invariant vacuum -- the
tree-level approximation for the no-boundary and tunneling cosmological
states. Rather it is a special state generating due to loop corrections
the peak-like distribution function which was obtained in \cite{qsi,qcr}
where it was shown to be drastically different from that of the DeSitter
vacuum. The contribution of this mode to radiation current, violating the
DeSitter invariance of effective equations, goes beyond the scope of this
paper. It is likely that by the big $|\xi|$ mechanism of the above type this
graviton-inflaton sector does not contribute to the leading order of
$1/|\xi|$-expansion, which justifies discarding its peculiarities.

With this reservation, the Wightman Green's function (\ref{3.5}) can be
regarded the Euclidean DeSitter invariant one. In Lorentzian DeSitter
spacetime it can be obtained by a proper analytic continuation from
the unique {\it regular} Green's function on the Euclidean section of
the DeSitter space \cite{Allen} -- a 4-dimensional sphere of the
radius $1/H(\varphi)$. Taken together with the local
Schwinger-DeWitt expansion of $J^{\rm rad}(x)$ discussed above
this means that the radiation current itself can be obtained by
this analytic continuation from the Euclidean radiation current which, in
its turn, expresses in terms of the Euclidean effective action
	\begin{eqnarray}
	&&J^{\rm rad}_{\rm E}(x)=\frac{\delta{\mbox{\boldmath $\Gamma$}}
	^{\rm loop}}{\delta\phi(x)},    \label{3.8}\\
	&&{\mbox{\boldmath $\Gamma$}}^{\rm loop}=
	\frac 12\,{\rm Tr}\,{\rm ln}\,\frac{\delta^2
	{\mbox{\boldmath $I$}}[\,\phi\,]}
	{\delta\phi\,\,\delta\phi}+...,       \label{3.9}
	\end{eqnarray}
where ${\mbox{\boldmath $I$}}[\,\phi\,]$ is the classical Euclidean action
of the theory related to the Lorentzian action by standard Wick rotation
	\begin{eqnarray}
	{\mbox{\boldmath $I$}}[\,\phi\,]=-iS[\,\phi\,]\,
	\Big|_{\,-+++\,\,\rightarrow\,\,++++}.   \label{3.10}
	\end{eqnarray}

Thus the Lorentzian effective equations in the approximation of
{\it local Schwinger-DeWitt expansion} can be obtained by analytically
continuing back to Lorentzian signature from the Euclidean effective
equations
	\begin{eqnarray}
	\frac{\delta{\mbox{\boldmath $\Gamma$}}}
	{\delta\phi(x)}=0                                 \label{3.11}
	\end{eqnarray}
where the Euclidean effective action
${\mbox{\boldmath $\Gamma$}}={\mbox{\boldmath $I$}}
+{\mbox{\boldmath $\Gamma$}}^{\rm loop}$ is calculated within this
local low-derivative expansion\footnote
{
This simple rule of obtaining the Lorentzian effective equations
from their Euclidean counterpart is an artifact
of two important properties: i) analytic relation between the Green's
functions on the Lorentzian and Euclidean sections of the DeSitter geometry
and ii) approximation of local low-derivative expansion. In the approximation
of rapidly varying fields nonlocal effective equations for {\it expectation
values} do not even have the form of the functional derivative of some
effective action \cite{beyond}. This is a distinctive feature of
the diagrammatic technique for expectation values which is different from
the conventional technique for matrix elements between two different
states -- in and out vacua.
} or the quasilocal expansion of \cite{Hu} (also considered in the
cosmological context in \cite{HuS,Hu1}). For the purposes of our slow-roll
approximation we need only the first three terms of this expansion in powers
of derivatives. They reproduce the structures of the classical gravitational
action (\ref{action1})
	\begin{eqnarray}
	{\mbox{\boldmath $\Gamma$}}[\,\varphi,g_{\mu\nu}]=
	\int d^{4}x g^{1/2}\left(V_{\rm eff}(\varphi)
	-U_{\rm eff}(\varphi)\,R +
	\frac{1}{2}G_{\rm eff}(\varphi)\,g^{\mu\nu}\varphi_{,\mu}
	\varphi_{,\nu}+...\right).                        \label{3.12}
	\end{eqnarray}
with the effective coefficient functions of the inflaton field
	\begin{eqnarray}
	&&V_{\rm eff}(\varphi)=V(\varphi)+V^{\rm loop}(\varphi), \nonumber\\
	&&G_{\rm eff}(\varphi)=1+G^{\rm loop}(\varphi),\nonumber\\
	&&U_{\rm eff}(\varphi)=U(\varphi)+U^{\rm loop}(\varphi),
	\end{eqnarray}
modified by quantum terms (the overall sign of this expression differs
from (\ref{action1}) in view of the Wick rotation to the Euclidean signature).
The latter will be built in the next section.

\section{One-loop effective action in the slow-roll
approximation: massive GUT sector}
\hspace{\parindent}
For massive fields the inverse propagator in (\ref{3.9}) generically has the
form of the covariant differential operator acting in the vector space
labeled by their isotopic indices
 	\begin{eqnarray}
	&&\frac{\delta^2
	{\mbox{\boldmath $I$}}[\,\phi\,]}
	{\delta\phi(x)\,\,\delta\phi(y)}=
	\left({\mbox{\boldmath$F$}}(\nabla)
	-m^2\hat{1}\right)\,\delta(x-y),            \label{4.1}\\
	&&{\mbox{\boldmath$F$}}
	=\Box+\hat{P}-\frac16\hat{1},\;\;\;\;
	\Box\equiv g^{\mu\nu}\nabla_\mu\nabla_\nu,        \label{4.2}
	\end{eqnarray}
with the spacetime-dependent potential term $\hat{P}-\hat{1}R/6$ (in
which the curvature scalar term is extracted for reasons of convenience).
The  matrix form of its coefficients is denoted by hats and $\hat{1}$ means
the unit matrix. The one-loop effective action of such fields
	\begin{eqnarray}
	\mbox{\boldmath$\SGamma$}^{\rm 1-loop}
	=\frac12\,{\rm Tr}\,\ln\,\left({\mbox{\boldmath$F$}}
	-m^2\,\hat{1}\right)                            \label{4.3}
	\end{eqnarray}
can be decomposed for large mass $m$ in the local Schwinger-DeWitt series
\cite{DW,BarvV}
	\begin{eqnarray}
	&&\mbox{\boldmath$\SGamma$}^{\rm 1-loop}=
	-\frac1{32\pi^2}\,\int d^4x\,g^{1/2}\,
	{\rm tr}\,\left\{\frac12\left(\frac1{2-\omega}-\ln\frac{m^2}{\mu^2}
	+\frac32\right)\,m^4\hat{a}_0(x,x)\right.\nonumber\\
	&&\quad\qquad+\left(\frac1{2-\omega}-\ln\frac{m^2}{\mu^2}
	+1\right)\,m^2\hat{a}_1(x,x)+
	\left(\frac1{2-\omega}-\ln\frac{m^2}{\mu^2}\right)\,
	\hat{a}_2(x,x)\nonumber\\
	&&\qquad\qquad\qquad\qquad\qquad\left.
	+\sum_{n=1}^{\infty}\frac{(n-1)!}{m^{2n}}\,
	\hat{a}_{n+2}(x,x)\,\right\},\;\;\;\omega\rightarrow 2. \label{4.4}
	\end{eqnarray}
Here $\hat{a}_n(x,x)$ are the Schwinger-DeWitt coefficients that can be
systematically calculated for generic theory as spacetime invariants of
growing power in spacetime and fibre bundle curvatures, potential term
of the operator and their covariant derivatives. For example
\cite{DW,Gilkey,BarvV,Avram}
	\begin{eqnarray}
	&&\hat{a}_0(x,x)=\hat{1},                    \label{4.5}\\
	&&\hat{a}_1(x,x)=\hat{P},                    \label{4.6} \\
	&&\hat{a}_2(x,x)=\frac1{180}(R_{\mu\nu\alpha\beta}^2-R_{\mu\nu}^2
	+\Box R)\hat{1}+\frac12\,\hat{P}^2
	+\frac1{12}\hat{\cal R}_{\mu\nu}^2
	+\frac16\,\Box\hat{P},                       \label{4.7}\\
	&&\hat{a}_3(x,x)=\frac1{12}(\,\nabla\hat{P})^2+
	\frac1{12}(\hat{P}\,\Box\hat{P}
	+\Box\hat{P}\,\hat{P}\,)+...,                  \label{4.7a}
	\end{eqnarray}
where $\hat{\cal R}_{\mu\nu}$ determines the commutator of covariant
derivatives acting on fields $\phi$,
	$(\nabla_\mu\nabla_\nu-\nabla_\nu\nabla_\mu)\,\phi =
	\hat{\cal R}_{\mu\nu}\,\phi$
and in $\hat{a}_3(x,x)$ only terms bilinear in $\hat{P}$ with two
derivatives are retained. In (\ref{4.4}) $\omega$ is half the dimensionality
of spacetime serving as a parameter of the dimensional regularization,
$\mu^2$ is a parameter reflecting the renormalization ambiguity and
${\rm tr}$ denotes the matrix (super)traces with respect to isotopic indices
of $\hat{a}_n(x,x)$.

In our case the only background field arguments of the effective action
consist of slowly varying scalar field and spacetime metric with the
curvature satisfying inequality (\ref{3.7}). From the structure of
equations (\ref{4.5})-(\ref{4.7}) it then follows that the renormalized
effective action
	\begin{eqnarray}
	&&\mbox{\boldmath$\SGamma$}^{\rm 1-loop}=
	\frac1{32\pi^2}\int d^4x\,g^{1/2}\,{\rm tr}\,\left\{\,
	\frac{m^4}2\left
	(\ln\frac{m^2}{\mu^2}\,\hat{1}-\frac32\right)\,
	\hat{1}\right.\nonumber\\
	&&\qquad\qquad\qquad\left.
	-m^2\,\left(\ln\frac{m^2}{\mu^2}-1\right)
	\,\hat{a}_1(x,x)+\ln\frac{m^2}{\mu^2}
	\;\hat{a}_2(x,x)
	+\frac{\hat{a}_3(x,x)}{m^2}\,\right\}+...     \label{4.8}
	\end{eqnarray}
is dominated by the contribution of the effective potential due to
$\hat{a}_0(x,x)=\hat{1}$ and the gravitational and inflaton kinetic terms
of second order in derivatives originating from $\hat{a}_1(x,x)$,
	\begin{eqnarray}
	\int d^4x\,g^{1/2}\,{\rm tr}\,\hat{a}_1(x,x)=
	\int d^4x\,g^{1/2}\,\left[\,u\,R
	+g\,(\nabla\varphi)^2\,\right]               \label{4.9}
	\end{eqnarray}
with some coefficients $u$ and $g$ depending on spin, $\hat{a}_2(x,x)$ and
$\hat{a}_3(x,x)$. For minimally coupled fields $g=0$, and $\hat{a}_2(x,x)$
gives only a fourth-order contribution of the derivative expansion. The
radiative corrections to the kinetic term of the scalar field originate
from the effective $\varphi(x)$-dependence of masses $m^2\sim\varphi^2(x)$,
and their contribution comes from the terms (\ref{4.7a}) of the
third DeWitt coefficient.

As mentioned above, due to Higgs mechanism in the slowly
varying inflaton field the GUT particles of the model (\ref{1.2}) acquire
slowly variable masses \cite{qsi,qcr}
	\begin{eqnarray}
	&&m^2=(m_\chi^2,\;m_A^2,\;m_\psi^2),            \label{4.10}\\
	&&m_\chi^2=\frac{\lambda\,\varphi^2(x)}2,\;\;
	m_A^2=g_A^2\,\varphi^2(x),\;\;
	m_\psi^2=f_\psi^2\,\varphi^2(x).               \label{4.11}
	\end{eqnarray}
This $\varphi$-dependence of masses converts the Schwinger-DeWitt series
(\ref{4.8}) into the quasilocal expansion of refs. \cite{Hu,HuS,Hu1} in which
the sums over contributions of such massive particles of spin 0, spin 1
and spin 1/2 have the form (taking into account the statistics encoded
in the supertrace operation tr)
	\begin{eqnarray}
	&&{\rm tr}\,\sum \frac{m^4}2\,\hat{1}
	=\frac{\lambda\varphi^4}4 \mbox{\boldmath$A$}, \label{4.12} \\
	&&{\rm tr}\,\sum \frac{m^4}2\,
	\ln\frac{m^2}{\mu^2}\,\hat{1}
	=\frac{\lambda\varphi^4}4\,
	\left(\mbox{\boldmath$A$}
	\ln\frac{\varphi^2}{\mu^2}
	-\mbox{\boldmath$B$}\right),                   \label{4.13}
	\end{eqnarray}
where $\mbox{\boldmath$A$}$ is given by (\ref{A}) and
	\begin{eqnarray}
	\mbox{\boldmath$B$}=-\frac{1}{2\lambda}
	\left(\sum_{\chi} \lambda_{\chi}^{2}\ln(\lambda_\chi/2)
	+16\sum_{A} g_{A}^{4}\ln g_{A}^{2}
	- 16\sum_{\psi}f_{\psi}^{4}
	\ln f_{\psi}^{2}\right).                      \label{4.15}
	\end{eqnarray}
Similarly, in view of the known expressions for $\hat{a}_1(x,x)$ of
scalar, vector and spinor fields \cite{DW,FrTs,BarvV}
	\begin{eqnarray}
	&&{\rm tr}\,\sum m^2\,\hat{a}_1(x,x)
	=\frac1{12}\mbox{\boldmath$C$} \varphi^2\,R,  \label{4.16}\\
	&&{\rm tr}\,\sum m^2\,
	\ln\frac{m^2}{\mu^2}\,\hat{a}_1(x,x)
	=\frac1{12}\,\left(\mbox{\boldmath$C$}
	\ln\frac{\varphi^2}{\mu^2}-\mbox{\boldmath$D$}\right)
	\varphi^2 R,                                   \label{4.17}\\
	&&\mbox{\boldmath$C$}=\sum_{\chi} \lambda_\chi
	-4\sum_{A} g_A^2+4\sum_{\psi}f_\psi^2,          \label{4.18}\\
	&&\mbox{\boldmath$D$}=
	\sum_{\chi} \lambda_\chi\ln(\lambda_\chi/2)
	-4\sum_{A} g_A^2\ln g_A^2
	+4\sum_{\psi}f_\psi^2 \ln f_\psi^2.             \label{4.19}
	\end{eqnarray}
Therefore, the one-loop contributions of massive GUT fields to the
effective coefficient functions of the potential and curvature terms
in (\ref{3.12}) equal
	\begin{eqnarray}
	&&V^{\rm 1-loop}(\varphi)=
	\frac{\lambda\varphi^4}{128\pi^2}
	\left(\mbox{\boldmath$A$}
	\ln\frac{\varphi^2}{\mu^2}-\mbox{\boldmath$B$}
	-\frac32\mbox{\boldmath$A$}\right),             \label{4.20}\\
	&&U^{\rm 1-loop}(\varphi)=
	\frac{\varphi^2}{384\pi^2}\left(\mbox{\boldmath$C$}
	\ln\frac{\varphi^2}{\mu^2}-\mbox{\boldmath$D$}
	-\mbox{\boldmath$C$}\right).                     \label{4.21}
	\end{eqnarray}

The first two DeWitt coefficients of these GUT fields do not contribute to
the kinetic term of the inflaton. The contribution to $G_{\rm eff}(\varphi)$
in (\ref{3.12}) comes from $\hat{a}_3(x,x)$ containing the terms (\ref{4.7a}).
The $\varphi(x)$-dependent masses generate spacetime dependent $\hat{P}$ with
$\nabla_\mu\hat{P}\sim -\nabla_\mu[m^2(x)]\sim-\nabla_\mu[\varphi^2(x)]$ and
via these terms lead to finite (renormalization unambiguous)
contribution\footnote
{A detailed derivation is based on splitting the total potential term of the
operator (\ref{4.1})-(\ref{4.2}) into an auxiliary strictly constant mass
parameter and a spacetime dependent part $\hat{P}$ with a subsequent
perturbation theory in $\hat{P}$.}
	\begin{eqnarray}
	&&\int d^4 x\,g^{1/2}{\rm tr}\,
	\sum \frac{\hat{a}_3(x,x)}{m^2}=-\frac16\mbox{\boldmath$E$}
	\int d^4 x\,g^{1/2} (\nabla\varphi)^2+...,         \label{4.22}\\
	&&\mbox{\boldmath$E$}=\sum_{\chi} \lambda_\chi
	+8\sum_{A} g_A^2-8\sum_{\psi}f_\psi^2,          \label{4.23}
	\end{eqnarray}
where ellipses denote the terms of the fourth order in derivatives. Thus,
$G^{\rm 1-loop}(\varphi)$ generated by this sector of the theory does not
contain renormalization ambiguious logarithms and equals
	\begin{eqnarray}
	G^{\rm 1-loop}(\varphi)=
	\frac16\,\frac{\mbox{\boldmath$E$}}{32\pi^2}.   \label{4.24}
	\end{eqnarray}

For massless fields or the fields not coupled to the inflaton the local
Schwinger-DeWitt
expansion does not work and their effective action should be calculated
within alternative approximation schemes, like $\zeta$-functional method
used in \cite{zeta,qcr}. Comparison of the obtained coefficient functions with
the results of \cite{qcr} shows the coincidence of the dominant logarithmic
contribution proportional to $m^4\ln(m^2/\mu^2)$. Typical masses of
particles not coupled to the inflaton are much lower in magnitude than
the masses (\ref{4.11}) at the maximum of the distribution function
(\ref{1.1}), and the ratio of their quantum corrections to those of
(\ref{4.20}) is $m^4/\lambda\varphi^4\mbox{\boldmath$A$}\sim
\lambda\mbox{\boldmath$A$}/(64\pi^2)^2\ll 1$ (here we assume that with
$\delta=O(1)$ it follows from eq.(\ref{delta}) that
$m^2\sim m_P^2(\lambda/8\pi|\xi|$).
Thus, such fields can be discarded relative to the distinguished sector
of (\ref{1.2}). The only exception is the graviton-inflaton sector of
the model which explicitly involves large parameter $|\xi|$ and, thus,
apriori can give a big contribution. In the next section we show that
it is actually suppressed by the powers of $1/|\xi|$.

\section{One-loop effective action: graviton-inflaton sector}
\hspace{\parindent}
Renormalization of ultraviolet divergences in the theory with the action
(\ref{action1}) was considered in \cite{renorm}. Its one-loop divergences
have the form
	\begin{eqnarray}
	&&{\mbox{\boldmath $\Gamma$}}_{\rm div}^{\rm 1-loop}
	[\,\varphi,g_{\mu\nu}]=\frac1{32\pi^2(2-\omega)}
	\int d^4x\,g^{1/2}\left\{5\frac{V^2}{U^2}-2U^2
	\left(\frac{\partial\bar{V}}{\partial\phi}\right)^2
	+\frac12 U^2\left(\frac{\partial^2\bar{V}}
	{\partial\phi^2}\right)^2\right.             \nonumber\\
	&&\qquad\qquad\qquad+\left[\,\left(\frac{45}2 U^{-3}(U')^2+
	U^{-2}G\right)V-13U^{-2}U'V' \right.         \nonumber\\
	&&\qquad\qquad\qquad\qquad\left.-\left(\frac{25}4U^{-1}(U')^2+2G+
	\frac12 U'\frac{d}{d\varphi}\right)
	\frac{\partial^2\bar{V}}{\partial\phi^2}\,\right]
	(\nabla\varphi)^2                              \nonumber\\
	&&\qquad\qquad\qquad-\left[\,\frac{13}{3}U^{-1}V+\frac16 U
	\frac{\partial^2\bar{V}}{\partial\phi^2}\,\right]\,R \nonumber\\
	&&\qquad\qquad\qquad\left.
	+\frac{43}{60}R_{\alpha\beta}^2+\frac1{40}R^2+
	O(R\nabla^2\varphi, \nabla^4\varphi)
	\vphantom{\left(\frac{\partial^2\bar{V}}
	{\partial\phi^2}\right)^2}\right\},           \label{5.1}
	\end{eqnarray}
where $O(R\nabla^2\varphi, \nabla^4\varphi)$ denotes the terms of the
overall fourth order in derivatives -- part of them linear in the curvature
times second derivatives of the scalar field and the rest with quartic
derivatives of the scalar field\footnote
{
Important correction is in order here. The equation (\ref{5.1}) above
was presented in ref.\cite{renorm} under the number (2.72) with the erroneous
coefficient 5/2 of the first term instead of the correct coefficient 5.
}. In contrast with the renormalization by
matter fields considered above, here the one-loop counterterm also
includes kinetic term of the scalar field. The coefficients of this
expression -- renormalizing the effective potential, kinetic and Einstein
terms -- are complicated functions of the classical quantities $V(\varphi),\,
G(\varphi),\, U(\varphi)$ (for generality $G(\varphi)$ is taken here
different from 1 also at the classical level). They involve these functions
and their derivatives with respect to $\varphi$ denoted by primes, as well
as auxiliary inflaton potential
	\begin{eqnarray}
	\bar{V}(\phi)=\left.\frac{V(\varphi)}
	 {U^2(\varphi)}\right|_{\varphi=\varphi(\phi)}  \label{5.2}
	\end{eqnarray}
in the auxiliary parametrization of the scalar field $\varphi=\varphi(\phi)$
defined by the differential equation\footnote
{
The new scalar field $\phi$ and its potential $\bar{V}(\phi)$ arise in the
Einstein frame of the action (\ref{action1}), in which there is no nonminimal
coupling of $\phi$ to the curvature. Together with the reparametrization of
the scalar field the transition to this frame includes the conformal
transformation of the metric \cite{renorm}.
}
	\begin{eqnarray}
	\left(\frac{d\phi}{d\varphi}\right)^2=
	U^{-2}(\varphi)\left[U(\varphi)G(\varphi)
	+3U'^2(\varphi)\right].                          \label{5.3}
	\end{eqnarray}
The derivatives of $\bar{V}(\phi)$ express as
	\begin{eqnarray}
	&&\frac{\partial\bar{V}}{\partial\phi}=
	\frac{UV'-2U'V}{U^2(UG+3U'^2)^{1/2}},              \label{5.4}\\
	&&\frac{\partial^2\bar{V}}{\partial\phi^2}=
	\frac1{(UG+3U'^2)^2}\,\left[12 U^{-2}(U')^4 V
	-9U^{-1}(U')^3 V'\right.\nonumber\\
	&&\qquad\qquad\qquad+3(U')^2 V''-3U'U''V'+5U^{-1}(U')^2 GV
	-2U''GV\nonumber\\
	&&\qquad\qquad\qquad\left.+UGV''-\frac72 U'GV'+U'G'V
	-\frac12 UG'V'\right],                             \label{5.5}
	\end{eqnarray}
which makes the algorithm (\ref{5.1}) explicit in terms of the original
coefficient functions of the classical Lagrangian. With the tree-level
expressions for the latter
	\begin{eqnarray}
	V=\frac{m^2\varphi^2}2+\frac{\lambda\varphi^4}4,\;\;\;
	G=1,\;\;\; U=\frac{m_P^2}{16\pi^2}+\frac12|\xi|\varphi^2,
	\end{eqnarray}
one has the cancellation of the leading powers of $\xi$ and $\varphi$
in (\ref{5.4}) and (\ref{5.5}) for large $|\xi|\gg 1$ (till the end of this
section we discard in $V(\varphi)$ the subleading term $m^2 \varphi^2/2$)
	\begin{eqnarray}
	\frac{\partial\bar{V}}{\partial\phi}=
	O(1/|\xi|^3),\;\;\;\;
	\frac{\partial^2\bar{V}}{\partial\phi^2}=
	O(1/|\xi|^3)
	\end{eqnarray}
and from (\ref{5.1}) obtains the leading behaviour for the divergent parts
of the one-loop coefficient functions in the graviton-inflaton effective
action
	\begin{eqnarray}
	&&V^{\rm 1-loop}_{\rm div}=\frac1{32\pi^2(2-\omega)}
	\left[\,\frac54 \frac{\lambda^2\varphi^4}{\xi^2}
	+O(1/|\xi|^3)\,\right],   \\
	&&G^{\rm 1-loop}_{\rm div}=\frac1{32\pi^2(2-\omega)}
	\left[\,-7\frac{\lambda}{|\xi|}
	+O(1/|\xi|^2)\,\right],   \\
	&&U^{\rm 1-loop}_{\rm div}=\frac1{32\pi^2(2-\omega)}
	\left[\,\frac{13}6\frac{\lambda\varphi^2}{|\xi|}
	+O(1/|\xi|^2)\,\right].
	\end{eqnarray}

Similarly to heavy fields of the GUT-matter sector, these divergences
allow one to estimate the order of magnitude in $|\xi|$ of the renormalized
effective functions originating from the graviton-inflaton sector of the
model\footnote
{
For heavy massive fields the divergences yield the dominant term
of $1/m^2$-expansion (\ref{4.4}) by a simple rule -- replacement of the pole
in spacetime dimensionality by the logarithm of mass,
$1/(2-\omega)\rightarrow-\ln(m^2/\mu^2)$. For massless fields this rule
is not correct, but the order of magnitude of the result is still
encoded in the expression for the residue at this pole.
}
	\begin{eqnarray}
	V^{\rm 1-loop}_{\rm grav-infl}=O(1/|\xi|^2),\;\;\;
	G^{\rm 1-loop}_{\rm grav-infl}=O(1/|\xi|),\;\;\;
	U^{\rm 1-loop}_{\rm grav-infl}=O(1/|\xi|).         \label{5.11}
	\end{eqnarray}
These contributions are much smaller than their matter counterparts
(\ref{4.20}), (\ref{4.21}) and (\ref{4.24}). Thus, the graviton-inflaton
sector in the model with a large nonminimal coupling gives a negligible
contribution. In particular, it generates a very small anomalous scaling
behaviour on the DeSitter instanton -- $\zeta(0)$ in the zeta-function
technique, coinciding with the pole part of the effective action in
the dimensional regularization. In the slow-roll approximation of a
constant inflaton field on the DeSitter geometry with
$R=2V/U\simeq\lambda\varphi^2/|\xi|, R_{\mu\nu}=Rg_{\mu\nu}/4$ it equals
	\begin{eqnarray}
	\zeta(0)=2{\mbox{\boldmath $\Gamma$}}_{\rm pole}^{\rm 1-loop}
	=-\frac{171}{10}+O(1/|\xi|),
	\end{eqnarray}
as compared to a very big anomalous scaling of GUT fields quadratic
in $|\xi|$ \cite{qsi,qcr}.

\section{Tunneling vs no-boundary quantum states}
\hspace{\parindent}
According to the discussion of Sect.3 effective equations in the slow roll
approximation are given by classical equations (\ref{equation5a})
- (\ref{equation5}) with classical coefficient functions replaced by their
effective counterparts modified by loop corrections (\ref{4.20}),
(\ref{4.21}) and (\ref{4.24}). In particular, the quantum rolling force
equals
	\begin{eqnarray}
	F_{\rm eff}(\varphi)=
	-\frac{U_{\rm eff}^3}{U_{\rm eff}+3U_{\rm eff}'^2}\,
	\frac d{d\varphi}\left(\frac{V_{\rm eff}}
	{U_{\rm eff}^2}\right).                          \label{6.1}
	\end{eqnarray}
These equations are the same for both quantum states -- no-boundary and
tunneling. Initial conditions are, however, different: initial value of
the mean inflaton field $\varphi_I$ yields the extremum of the distribution
function\footnote
{
In the model with large $|\xi|$ the local extremum of the
distribution function is very sharp, so that $\varphi_I$ in the leading order
coincides with the point of this extremum.
} which is different for these two states (\ref{1.1}). Thus, to
sort out the
nature of quantum evolution from this probability peak, that is to find the
sign of the rolling force at $\varphi_I$, $F_{\rm eff}(\varphi_I)$, we have
to consider these two states separately. Consider first the no-boundary case.

In the no-boundary distribution function (\ref{1.1}) the exponential is
determined by the {\em sum} of the classical Euclidean action on the
DeSitter instanton
	\begin{eqnarray}
	\mbox{\boldmath$I$}(\varphi)=
	\int\limits_{\bf\rm DS} d^4x\,g^{1/2}
	\left.\Big(-U(\varphi)R+V(\varphi)\,\Big)\,
	\right|_{\,\varphi={\rm const}}=
	-\frac{96\pi^2 U^2(\varphi)}{V(\varphi)}      \label{6.2}
	\end{eqnarray}
and the one-loop effective action which in the slow-roll approximation
on the DeSitter sphere of radius $H^{-1}=[\,6U/V\,]^{1/2}$ coincides with
the first order variation of the above expression under the variations
$\delta U=U^{\rm 1-loop}$ and $\delta V=V^{\rm 1-loop}$
	\begin{eqnarray}
	&&{\mbox{\boldmath $\Gamma$}}^{\rm 1-loop}(\varphi)
	=\int\limits_{\bf\rm DS} d^4x\,g^{1/2}
	\left.\Big(-U^{\rm 1-loop}(\varphi)R
	+V^{\rm 1-loop}(\varphi)\,\Big)\,
	\right|_{\,\varphi={\rm const}}   \nonumber\\
	&&\qquad\qquad\qquad=96\pi^2 \left(\,\frac{U^2(\varphi)}
	{V^2(\varphi)}\,V^{\rm 1-loop}(\varphi)
	-\frac{2U(\varphi)}
	{V(\varphi)}\,U^{\rm 1-loop}(\varphi)\right).      \label{6.3}
	\end{eqnarray}
Thus up to higher order terms in powers of quantum corrections the exponential
of the no-boundary distribution function coincides with on-shell effective
action on the effective DeSitter spacetime of the radius
$H_{\rm eff}^{-1}=[\,6U_{\rm eff}/V_{\rm eff}\,]^{1/2}$
	\begin{eqnarray}
	{\mbox{\boldmath $\Gamma$}}(\varphi)=
	\mbox{\boldmath$I$}(\varphi)+
	{\mbox{\boldmath $\Gamma$}}^{\rm 1-loop}(\varphi)=
	-\frac{96\pi^2\,[\,U_{\rm eff}(\varphi)\,]^2}
	{V_{\rm eff}(\varphi)}
	+O(\hbar^2).                                \label{6.4}
	\end{eqnarray}
Here $\hbar$ symbolically denotes the quantum terms proportional to either
of the following one-loop combinations of coupling constants of Sect. 4,
$\hbar=(1/32\pi^2)(\mbox{\boldmath$A$},\mbox{\boldmath$B$},\mbox{\boldmath$C$},
\mbox{\boldmath$D$},\mbox{\boldmath$E$})$. Therefore, in the domain of the
slow roll approximation the no-boundary distribution function equals
	\begin{eqnarray}
	\rho_{\rm NB}(\varphi)={\rm const}\,\exp\left
	\{\frac{96\pi^2\,[\,U_{\rm eff}(\varphi)\,]^2}
	{V_{\rm eff}(\varphi)}+O(\hbar^2)\right\}.    \label{6.5}
	\end{eqnarray}
Note that this expression is valid irrespective of the concrete form of
radiative corrections $U^{\rm loop}$ and $V^{\rm loop}$, and $O(\hbar^2)$
here can be regarded as a contribution of multiloop orders.

Now, comparing this expression with (\ref{6.1}) one can see that in the
approximation of the above type the quantum rolling force for the mean
inflaton field at its probability maximum is identically zero
	\begin{eqnarray}
	F_{\rm eff}^{\rm NB}(\varphi_I)\sim
	\left.\frac d{d\varphi}\ln\rho_{\rm NB}(\varphi)\,
	\right|_{\,\varphi=\varphi_I}\equiv 0.       \label{6.6}
	\end{eqnarray}
This one-loop result holds for any model of matter sources and any structure
of their quantum corrections. It implies that the no-boundary model of
quantum origin of inflation does not satisfy the requirement of a finite
duration of inflation stage: the quantum part of the rolling force cancels
its classical part to zero but not reverses its
sign to provide the decrease of the inflaton field and gradual exit from
inflation (remember that with classical equations of motion the rolling force
was of a wrong -- positive -- sign).

Let us go over to the tunnelling quantum state of the Universe. In contrast
with the no-boundary case the exponential of the distribution
function contains the {\it difference} of the classical Euclidean action and
the one-loop effective action. This difference for large $|\xi|\gg 1$
generates a probability peak with
the field (\ref{2.1}) in the opposite range of the parameter (\ref{delta}),
$\delta>-1$. According to the detailed discussion of \cite{qsi,qcr}
comparison of this peak with the observational restrictions on the
inflationary scenario -- the lower bound of the classically induced e-folding
number (\ref{2.11}), $N>60$ -- leades to the estimate on the universal
combination of coupling constants $\mbox{\boldmath$A$}$ \cite{qsi,qcr},
$\mbox{\boldmath$A$}=O(1)$ and similar estimates for the other combinations
$(\mbox{\boldmath$B$},\mbox{\boldmath$C$},
\mbox{\boldmath$D$},\mbox{\boldmath$E$})=O(1)$ (because such mechanisms
as supersymmetry that could provide cancellation of separately big terms of
(\ref{A}) do not seem to apply in our model \cite{Miele}). Therefore,
all the quantities that were symbolically denoted above by $\hbar$
comprise very small parameters $\mbox{\boldmath$A$}/32\pi^2\ll 1,\;
\mbox{\boldmath$B$}/32\pi^2\ll 1$, etc. With these bounds and in the limit
of large nonminimal coupling $|\xi|$ one easily finds the tunneling
rolling force by substituting the one-loop expressions of Sect.4 to
(\ref{6.1}). In the vicinity of the tunneling maximum $\varphi_I$ the result
reads
	\begin{eqnarray}
	F^{\rm T}_{\rm eff}(\varphi)=-\frac{\lambda m_P^2
	(1+\delta)}{48\pi\xi^2}\,\varphi
	\left(1+\frac{\varphi^2}{\varphi^2_I}\right)
	+O(1/|\xi|^3), \,\,\,\,\,1+\delta>0.         \label{6.7}
	\end{eqnarray}
The second term here is the one-loop quantum contribution which obvioulsly
doubles the classical force (\ref{2.2}) at $\varphi=\varphi_I$. Thus it
does not qualitatively change the predictions of classical equations of
motion: it guarantees the slow decrease of the inflaton field during the
inflation and results in its finite duration with slightly different
e-folding number
	\begin{eqnarray}
	N\simeq -3\int\limits_0^{\varphi_I}
	d\varphi\,\frac{[\,H_{\rm eff}(\varphi)\,]^2}
	{F^{\rm T}_{\rm eff}(\varphi)}
	\simeq \frac{48\pi^2}{\mbox{\boldmath $A$}}\,\ln 2\,
	\left[\,1+O(\varepsilon\ln\varepsilon)\,\right].   \label{6.8}
	\end{eqnarray}
Similarly to eq.(\ref{2.11}) (see the corresponding footnote), we discard
here subleading terms coming from subdominant quantum corrections and
lower order terms of $U_{\rm eff}(\varphi)$ and $V_{\rm eff}(\varphi)$
dominating at later stages of inflation, $\varepsilon=
(1/32\pi^2)(\mbox{\boldmath$A$},\mbox{\boldmath$B$},\mbox{\boldmath$C$},
\mbox{\boldmath$D$},\mbox{\boldmath$E$})\ll 1$. Thus, the restriction on
the minimal duration of inflation $N\geq 60$ at the quantum level does not
qualitatively change the classical bound $(\mbox{\boldmath $A$}\leq 7.9)$
	\begin{eqnarray}
	\mbox{\boldmath $A$}\leq 5.5.                  \label{6.9}
	\end{eqnarray}

For completeness let us also present the expression for the
rolling force in the vicinity of the probability maximum in the case of
the no-boundary state. It can be obtained from eq.(\ref{6.7}) by
inverting the sign of the second term
	\begin{eqnarray}
	F^{\rm N\!B}_{\rm eff}(\varphi)=-\frac{\lambda m_P^2
	(1+\delta)}{48\pi\xi^2}\,\varphi
	\left(1-\frac{\varphi^2}{\varphi^2_I}\right)
	+O(1/|\xi|^3), \,\,\,\,\,1+\delta<0.              \label{6.10}
	\end{eqnarray}
In accordance with (\ref{6.6}) it vanishes at $\varphi_I$ and also changes
the sign here, which means the stability of this point. This situation
is completely similar to the tree-level situation \cite{Vilenkin} -- the
Lorentzian DeSitter Universe nucleates from the gravitational (half)instanton
in the minimum of the effective potential
$V_{\rm eff}(\varphi)/U^2_{\rm eff}(\varphi)$ and stays there with a constant
stable value of the inflaton field.

\section{Conclusions and discussion}
\hspace{\parindent}
The above results add a number of new facets to the dilemma of two
wavefunctions -- tunneling and no-boundary ones -- as candidates for the
quantum state of the early Universe \cite{Vilenkin,tvsnb,Vil-pol,Bousso}
capable of generating via inflationary
scenario the observable large scale structure. They confirm our
previous conclusion \cite{qsi,qcr} that the tunneling state is more likely to
describe its quantum cosmological origin: it predicts a finite
inflationary epoch characterized by the e-folding number (\ref{6.8})
and the rolling force (and $\dot\varphi\neq 0$) nonvanishing from the outset
of inflation. This guarantees the smallness of density perturbations
inverse proportional to $\dot\varphi$, which can be important for the
inflation models with $\Omega\neq 1$ \cite{LindeM,open}.
The finiteness of inflation stage is a result of a weak breakdown of the
DeSitter invariance existing already at the classical level -- slow roll
of the inflaton field down the hill of the inflaton potential. In the
tunneling case one-loop quantum corrections enhance this breakdown and
decrease the classical e-folding number by a factor of $\ln 2$ in (\ref{6.8}).

In the case of the no-boundary quantum state the situation is different.
Classically (that is in {\em classical} equations of motion) at the maximum
of the one-loop distribution function the sign of the deviation from DeSitter
invariance is opposite -- the inflaton potential has a negative slope.
One-loop quantum corrections in {\em effective}
equations exactly compensate this deviation, so that the mean inflaton field
remains constant in a stable minimum of the effective one-loop potential.
This conclusion is universal for it does not depend on the form of one-loop
corrections. This universality follows from the fact that, unlike the
tunneling case, for the no-boundary state one and the same dynamical
principle is laid in the foundation of quantum initial conditions
(the distribution function $\rho_{\rm NB}(\varphi)$) and effective
equations for mean fields
-- the path integral formulation of the no-boundary wavefunction \cite{HH,H}.
Therefore, one and the same effective inflaton potential enters both the
effective equations and probability distribution, which makes the rolling
force vanishing in a stable point of the maximum probability. Thus, the
no-boundary quantum state is intrinsically more symmetric (in the DeSitter
sense) than the tunneling one. This makes it unsatisfactory from the
viewpoint of applications in the theory of the early universe\footnote
{
Infinite duration of inflation with constant inflaton, $\dot\varphi\simeq 0$,
and generation of exceedingly large perturbations (inverse proportional to
$\dot\varphi$) contradict a generally accepted inflationary scenario with
the observable magnitude of microwave backround radiation anisotropy,
$\Delta T/T\sim 10^{-5}$.
}, but renders it very attractive in other problems needing DeSitter invariant
vacuum. The exception from this rule exists in the class of models considered
in the tree-level approximation \cite{khalat1,khalat2,khalat} which can also
be ascribed to the no-boundary class\footnote
{
Only by the sign of the exponentiated Euclidean action in the underbarrier
domain. From the topological viewpoint this is not a no-boundary state,
because it is more likely to describe the underbarrier penetration
from one classically allowed region (with small scale factor) to
another one with the scale factor infinitely growing to infinity.
}.
In these models with complex inflaton field the probabilistically preferred
solutions can describe models with finite inflation stage due to the
presence in the inflaton potential of the centrifugal term produced by
conserved isotopic charge. For nonminimal inflaton field in these models
one can get probability peaks for both no-boundary
and tunneling wave functions by means of a proper choice of parameters
\cite{khalat2,khalat}.

The preferred nature of the tunneling wavefunction of the above type
sounds coherent with recent conclusions of Linde \cite{Lindop} who extended
the Hawking-Turok no-boundary mechanism of creating the {\it open}
universe \cite{HawkTur} to the tunneling case. In principle, our results
obtained for the closed model can be directly extended to open models
by using the Hawking-Turok instanton. This instanton is singular, but
according to their calculations \cite{HawkTur} this singularity is mild
enough to retain finite and practically the same value of the instanton action
as in (\ref{6.2}). It is likely that the same will be true also for the
anomalous scaling of loop corrections, so that our mechanism of generating
the probability peak at GUT scale will also work in the open
Universe \cite{Barvop}\footnote
{In contrast with initial conditions, the problem of effective equations for
the Hawking-Turok model of open inflation is much harder, because the local
Shwinger-DeWitt expansion and slow roll approximation break down near the
singularity of the Hawking-Turok instanton \cite{Barvop}.
}.
Then, no anthropic principle -- an obvious retreat for theory -- should be
invoked to reach the conclusions of \cite{HawkTur} and improve them by finding
the value of $\Omega$ satisfying $1>\Omega\gg 0.01$. A similar improvement
looks true regarding the implications of tunnelling wavefunction for open
inflation by Linde \cite{Lindop}. To find $\Omega$ in this case one should
not appeal to special supergravity induced potentials generating
inflation only in the limited range of $\varphi$ \cite{Lindop} (which by
itself can be regarded as a mild form of the anthropic principle), but should
calculate it from the obtained probability maximum. The GUT scale of this
peak, in particular, rules out the necessity of considering such potentials
which obviously manifest themselves at the much higher supergravitational
energy scale.

Here a natural question arises, how seriously should be considered these
predictions based only on the one-loop approximation. As it was discussed
in \cite{qcr}, the main justification of the semiclassical expansion comes
from the fact that the energy scale of the system (\ref{2.1}) is suppressed
relative to the Planck scale by the numerical factor $\sqrt{\lambda}/|\xi|$.
In the model with nonminimal inflaton \cite{nonmin} this factor determines
the magnitude of microwave background anisotropy which is well known from
the COBE \cite{COBE} and Relict \cite{Relict} satellite experiments at the
level of $10^{-5}$. This
leads to the large value of $|\xi|$ and, as a consequence, to the large
value of the effective Planck mass (\ref{3.6}) and suppression of higher
powers of curvatures (\ref{3.7}) characteristic of multi-loop contributions.
The smallness of the graviton-inflaton contribution (\ref{5.11}), in
particular,
is a direct consequence of such an improvement of the semiclassical expansion
due to the replacement of $1/m_P^2$ by $1/(m_P^2+8\pi|\xi|\varphi^2)$.

In the GUT sector of the model the smallness of multi-loop orders can be
regarded as a consequence of the bound (\ref{6.9}) and its
corollaries\footnote
{
As discussed in \cite{qcr}, except rather improbable mechanisms of
cancellation
between different terms of (\ref{A}) the smallness of $\mbox{\boldmath$A$}$
is only guaranteed by small values of all the couplings
$\lambda_\chi,g_A,f_\psi$ (as is the case with running gauge coupling
constants at the grand unification point). This guarantees small values of
other universal combinations of couplings $\mbox{\boldmath$B$},
\mbox{\boldmath$C$},\mbox{\boldmath$D$},\mbox{\boldmath$E$}$.
}
	\begin{eqnarray}
	\varepsilon \equiv\frac1{32\pi^2}
	(\mbox{\boldmath$A$},\mbox{\boldmath$B$},\mbox{\boldmath$C$},
	\mbox{\boldmath$D$},\mbox{\boldmath$E$})\ll 1.
	\end{eqnarray}
The multi-loop orders roughly proportional to powers of
$\varepsilon$ are thus small. In connection with
this observation it is worth explaining the following paradox. The smallness
of $\varepsilon$ obviously implies the smallness of all quantum effects in the
system including the one-loop order, so how could the latter qualitatively
change the predictions of the tree-level theory -- replace a flat graph
of the tree-level distribution by the probabilty peak advocated in
\cite{qsi,qcr}? The explanation is as follows. The smallness of $\varepsilon$
indeed leads to
small quantum terms of the effective {\em Lagrangian}. Large quantum
effects of the effective {\em action} originate from multiplying the
Lagrangian by the 4-volume of the DeSitter instanton, $8\pi^2/3H^4(\varphi)$,
which brings to life a large factor of $|\xi|^2$. The classical part of the
action, exactly by the same mechanism, is also quadratic in $|\xi|$, but its
$\varphi$-dependent part is at most {\em linear} in $|\xi|$ \cite{qsi,qcr}.
This allows one to balance this classical part by the quantum term
($\sim\varepsilon|\xi|^2$) and reach a nontrivial maximum of the
probability distribution even though $\varepsilon\ll 1$.

All these arguments have certainly a
qualitative nature and remain valid only for fields in the vicinity of
the obtained probability maximum. A rigorous proof would require a careful
consideration and bookkeeping of multi-loop orders and the nonperturbative
analysis of the $\varphi\rightarrow\infty$ limit. At present, however,
we have a reliable algorithm for the probability distribution (\ref{1.1}) only
in the one-loop approximation \cite{tunnel}. Its extension to higher orders
has not yet been done and might be rather nontrivial in view of
intrinsic problems of the Dirac quantization of constrained systems
\cite{BKr,geom}\footnote
{
The starting point for the derivation of the probability algorithm is the
solution of the Wheeler-DeWitt equations -- quantum Dirac constraints in
the coordinate representation of the canonical commutation relations. The
physical inner product (generating the probability amplitudes) in this
representation is known only in the one-loop approximation
\cite{BKr,BarvU,geom} -- the measure of this inner product apparently has
essentially perturbative nature. Thus, the multi-loop extension of the
algorithm for probability distribution requires developing the fundamental
aspects of quantization formalism for constrained dynamical systems.
}.
Therefore, regarding the nonperturbative behaviour we can only put forward
a number of hypotheses.

It is likely that for the no-boundary state the exact probability distribution
is given by the full Euclidean effective action including all loop
corrections. Then in the slow roll approximation it coincides with the
expression (\ref{6.5}) involving full coefficient functions
$U_{\rm eff}(\varphi)$ and $V_{\rm eff}(\varphi)$ without extra $O(\hbar^2)$
terms. Then the high-energy limit of $\rho_{\rm NB}(\varphi)$ at
$\varphi\rightarrow\infty$ is encoded in the corresponding behaviour of
the ratio of these functions $U^2_{\rm eff}(\varphi)/V_{\rm eff}(\varphi)$.
Their {\it hypothetic} asymptotic behaviour can be motivated by the one-loop
approximation (\ref{4.20}) - (\ref{4.21}) with some {\rm effective}
constants $\mbox{\boldmath$A$}_{\rm eff}$ and $\mbox{\boldmath$C$}_{\rm eff}$
including all multi-loop corrections\footnote
{The generalized renormalization group in the model of graviton nonminimally
coupled to inflaton considered in our work \cite{renorm} leads to
the different asymptotics for $V_{\rm eff}(\varphi)$,
$V_{\rm eff}(\varphi)\sim \varphi^4\left[\ln(\varphi^2/\mu^2)\right]^2$,
but this behaviour
is apparently based on an unapproapriate branch of renormalization flow
starting from the over-Planckian conformally invariant phase, which does not
interpolate between the latter and the GUT phase.
}. Then the hypothetical high-energy
behaviour of the {\it no-boundary} distribution function looks like
	\begin{eqnarray}
	\rho_{\rm NB}(\varphi)\sim\exp
	\left\{\frac{\,\mbox{\boldmath$C$}^2_{\rm eff}}
	{6\lambda\mbox{\boldmath$A$}_{\rm eff}}\,\ln\varphi\right\},
	\;\;\;\;\varphi\rightarrow\infty,
	\end{eqnarray}
and its normalizability (and the requirement of suppression of the high
energy scales) imposes the restriction on the effective constant
$\mbox{\boldmath$A$}_{\rm eff}$, $\mbox{\boldmath$A$}_{\rm eff}<0$, quite
opposite to the one obtained in our earlier work on the one-loop
normalizability of the no-boundary wavefunction \cite{norm}. This
contradiction should, however, be regarded with a big share of criticism,
because it involves too many hypotheses. As regards the tunneling
quantum state in the nonperturbative regime, we can say even less than
about the no-boundary one. In the one-loop approximation quantum and
classical terms enter the game with different signs, so it is
tempting to conjecture that the exact answer can again be obtained from
semiclassically expanded full effective action by formally inverting the
sign of the Planck mass squared, but this conjecture demands a careful check.

Another limitation of the obtained results regards the contribution of the
quantum mechanical sector of the homogeneous inflaton mode. This mode, as
mentioned in Sect.3, is effectively massless and, thus, does not possess
the DeSitter invariant vacuum \cite{Allen} -- the problem which in the
quantum cosmological context implies the absence of normalizable no-boundary
and tunneling states considered in the tree level approximation. The main
achievement of refs.\cite{qsi,qcr} is that beyond this approximation in the
model with large $|\xi|$ this problem can have a solution in the form of
a sharply peaked distribution with the parameters of the peak
(\ref{2.1}) - (\ref{peak}). Therefore, the Green's function of this mode
(\ref{3.5}), participating in the radiation current of the effective
equations, should be defined relative to such a peaked state. Apriori,
it is not DeSitter invariant and its contribution to Lorentzian effective
equations even in the slow roll approximation does not follow from the
Euclidean effective action by Wick rotation (\ref{3.10}) - (\ref{3.11}).
The method of calculating this contribution is currently under study -- it
involves actual Hamiltonian reduction of the system to the explicit
physical degree of freedom in the homogeneous gravity-inflaton sector of
the model and raises important gauge dependence issues\footnote
{
The problem of the off-shell gauge dependence of the effective gravitational
equations of motion \cite{Unique,BarvV} did not arise above, because this
gauge dependence (in the natural class of gauges fixing the
diffeomorphism invariance) affects only the graviton-inflaton sector of the
theory. The latter was discarded because its contribution is suppressed for
large nonminimal coupling $|\xi|$.
}. Despite this omission in the effective equations of quantum motion, there
is a strong believe that for large $|\xi|$ (that is, in the leading order of
the slow roll approximation) this does not affect our conclusions, because
all effects of the DeSitter invariance violation, including the
contribution of this quantum mechanical sector, are expected to belong to
the subleading order of $1/|\xi|$ and $\dot\varphi$ expansion. This, however,
requires a careful analysis which we postpone till later publication.

The slow roll approximation was a very important ingredient of all our
considerations. Being the attribute of the physical setting at the initial
stage of inflation it simultaneously served as a means to avoid the problem
of nonrenormalizability in local gravity theory: discarding (negligible)
higher powers of the curvature in local terms of the action leaves us with
the renormalization of only two generalized coupling constants --
cosmological term (effective potential $V_{\rm eff}$) and the gravitational
``constant'' $16\pi U_{\rm eff}$. Thus, approximately this brings us to
the class of perturbatively renormalizable theories -- the so-called
renormalization at the threshold \cite{FVil}. This approximation,
however, does not incorporate nonlocal quantum effects that correspond to
infinite summation of derivatives in the local Schwinger-DeWitt series
\cite{beyond}. Nonlocal terms of the effective equations for expectation
values cannot be obtained by the variational procedure with some nonlocal
effective action (see the footnote for eq.(\ref{3.11})). These terms
include the dissipation effects \cite{BeiLok} caused by particle creation
in the external (mean) field and become important in the end of inflation
at the reheating stage of the cosmological evolution \cite{alstar,espos}.
Their analysis should bring us to the complete cosmological scenario
including the formation of the observable large scale structure and
give the answer if the latter bears the imprint of the quantum
cosmological origin from the tunneling or no-boundary quantum states of
the Universe.

\section*{Acknowledgements}
\hspace{\parindent}
We are grateful to I.M.Khalatnikov, A.A.Starobinsky and A.V.Toporensky for
useful discussions. We also want to express our special thanks to B.Allen
and B.L.Hu for helpful discussions on the analytic extension properties in
the DeSitter spacetime. Helpful correspondence with A.Linde is also
deeply acknowledged. The work of A.O.B. was supported
by the Russian Foundation for Basic Research under grants No 96-02-16287 and
No 96-02-16295. A.Yu.K. is grateful for support by the Russian Foundation
for Basic Research under grants No 96-02-16220 and No 96-02-16287 and to
the grant of support of leading scientific schools 96-15-96458. This
work has also been supported in part by the Russian Research program
``Cosmomicrophysics''.

\end{document}